\documentclass{aip-cp}

\usepackage[numbers]{natbib}
\usepackage{rotating}
\usepackage{graphicx}

\begin{document}

\title{Weak Neutral Current Studies with Positrons}

\author[aff1]{Seamus Riordan\corref{cor1}}

\affil[aff1]{Argonne National Laboratory\\Argonne IL, USA 60439}
\corresp[cor1]{Corresponding author: sriordan@anl.gov}

\maketitle

\begin{abstract}
    Weak neutral current interactions with charged leptons have offered unique
    opportunities to study novel aspects of hadronic structure and search for
    physics beyond the standard model.  These studies in the medium energy
    community have been primarily through parity-violating processes with electron
    beams, but with the possibility of polarized positron beams, new and
    complementary observables can be considered in experiments analogous to
    their electron counterparts. Such studies include elastic proton,
    deep inelastic, and electron target scattering.  Potential
    positron neutral current experiments along with their
    potential physics reach, requirements, and feasibility are presented.
\end{abstract}

\section{Introduction and Formalism}
Weak neutral current studies with electron beams have been a powerful method to study unique properties of nucleons and nuclei
as well as provide a method to search for new physics outside of the standard model.  At low momentum transfers, an exchanged
neutral $Z$ boson will interfere with a virtual photon producing a parity-violating violating observable which can be separated
from the parity-conserving electromagnetic interaction.  The measured quantity is
typically a parity violating asymmetry $A_\mathrm{PV}$ generated by the two helicities of a lepton beam on an unpolarized target
taking a form
\begin{equation}
    \frac{\sigma_R - \sigma_L}{\sigma_R + \sigma_L} = A_\mathrm{PV} \propto \frac{G_F Q^2}{\sqrt{2} \pi \alpha} \times ...
\end{equation}
where $\sigma_{R,L}$ are the right and left-handed lepton cross section, $G_F$ is the Fermi coupling constant, $Q^2$ is the negative
four-momentum transfer squared, and $\alpha$ is the electromagnetic fine structure constant.  The ellipses represent the terms
which carry the details and structure of a particular interaction.  This asymmetry at the momentum transfers available with modern
electron beams on fixed targets is typically in the range of $10^{-6}$ to $10^{-3}$, which requires significant experimental control to observe.

From the unique couplings in the interference, these asymmetries provide measurements of the standard model couplings involving $\sin^2\theta_W$ \citep{qweak, e158}, charge-symmetry breaking contributions to nucleon elastic form factors, e.g. \citep{sample, g0, happex,happexiii}, and the neutron distributions within nuclei \citep{prex}.  From general helicity considerations, the parity violating asymmetries from an unpolarized target can be divided into forward and backwards components
\begin{equation}
            A  =  \frac{\sigma_R-\sigma_L}{\sigma_R+\sigma_L} \approx \frac{G_F Q^2}{\sqrt{2} \pi \alpha} \left[  D_\mathrm{f}(\theta) g^e_A g_V^\mathrm{target} + D_\mathrm{b}(\theta) g^e_V g_A^\mathrm{target} \right] 
\end{equation}
where $D_\mathrm{f,b}$ are forward and backwards components depending on the interaction and $g_A$ and $g_V$ are the vector and axial couplings in the neutral current weak interaction.  The vector coupling of the electron is $g_V^e = -\frac{1}{2} + 2 \sin^2\theta_W~\approx~0.04$ which is small by the value nature has chosen for the weak mixing angle, $\theta_W$.  This intrinsically suppresses the backwards part of the asymmetry which contains the axial structure of the target and are typically less well known than the vector counterparts.  This also must be done using kinematic separation of the forward and backwards parts requiring multiple measurements in different experimental configurations.  These axial components are also less constrained due to the fact that axial currents are not conserved.

With the additional consideration of positron beams, new possibilities become available.  Under charge conjugation of the leptonic vertex, the overall structure of the $\gamma-Z$ interference is effectively modified by the electron axial coupling undergoing a sign change \citep{PhysRevD.9.2171}.  In principle the positron parity-violating asymmetry contains similar levels of information, so the aforementioned studies by themselves do not dramatically change the experimental reach.  However, when combinations of both electron and positron data are made, the methods of extracting specific quantities changes and new observables become available.  The electron vector-target axial component can be extracted in the sum of the two parity violating asymmetries
\begin{equation}
    A^+_\mathrm{PV} - A^-_\mathrm{PV} = \frac{\sqrt{2} G_F Q^2}{\pi \alpha} g_V^e G_A^\mathrm{target}
\end{equation}
with $G_A$ containing the general axial structure of the interaction.  This object is the same size as the individual parity violating asymmetry contribution and subject to the same experimental issues, but at tree level provides a method of extraction that does not require kinematic separation of the forward and backwards parts.  Decades of effort have been put into controlling the production of electron beam suitable for parity-violating measurements would need to be reexamined given a different polarized positron production method.  Aside from this very serious issue, the reduced available current and polarization puts severe restrictions on possible measurements.  For a parity violating measurement, the figure of merit is proportional to $N P^2$ where $N$ is the detected number of counts in a process and $P$ is the beam polarization.  Given a typical electron beam parity violating experiment with 60~$\mathrm{\mu A}$ beam current and 85\% polarization, the approximate relative figure of merit for positron parity violating measurements using 100~nA beam with $60\%$ polarization is $10^{-3}$, i.e. requires 1000 times longer experiments, or absolute statistical error bars which are 30 times larger with equivalent measurement time.

The unpolarized charge asymmetry gives access to new non-parity violating axial-axial couplings which are not suppressed
\begin{equation}
    \frac{\sigma^+ - \sigma^{-}}{\sigma^+ + \sigma^{-}} = \frac{G_F Q^2}{\sqrt{2} \pi \alpha} g_A^e G_A^\mathrm{target}
\end{equation}
where $\sigma^+$ and $\sigma^-$ refers to the positron or electron cross section, respectively.

The extreme experimental challenge of this quantity is the charge normalization between electrons and positrons as this asymmetry is similar in scale to the parity violating asymmetry.  For example in the parity violating asymmetries, rapid flipping of the lepton helicity and feedback mechanisms are required to control for slow drifts in the beam properties which may introduce false asymmetries.  An analogous method would be to rapidly flip between charge states, which is technically difficult to do and would have to be done minimizing the change of other properties of the beam.  The advantage to this technique is higher available beam current when no polarization is required and a larger asymmetry containing the axial part.  Despite these issues, we consider several experimental scenarios.

\section{Potential Experiments and Reach}

\subsection{Elastic Proton Scattering}

Elastic scattering of the proton using the electroweak interaction gives information about that static structure of the proton and is invaluable to the understanding of the strong nuclear force.  In particular, by considering electromagnetic and weak processes together, tests of the standard model can be devised and nucleon properties such as charge symmetry violating contributions (e.g. strange quark contributions) and the axial charge can be measured. The parity-violating asymmetry at the Born level is given by \citep{doi:10.1146/annurev-nucl-102010-130419}
\begin{equation}
    A = \left[ \frac{-G_F Q^2}{4\pi \alpha \sqrt{2}} \right] \left[ \frac{ \epsilon G_E^\gamma { G_E^\mathrm{Z}} + \tau G_M^\gamma G_M^\mathrm{Z} + 2 g_V^e \epsilon' G_M^\gamma G_A^\mathrm{Z} }{ \epsilon \left( G_E^\gamma \right)^2 + \tau \left( G_M^\gamma \right)^2 } \right]  
\end{equation}
where $G_{E,M}^{\gamma}$ are the electric and magnetic Sachs form factors, $G_A^\mathrm{Z}$ is the axial form factor of the proton, $G_{E,M}^\mathrm{Z}$ are the vector form factors which couple to the $Z$,  $\epsilon$ is the virtual photon transverse polarization
\begin{equation}
    \epsilon = \left[ 1 + 2(1+\tau)\tan^2\frac{\theta}{2} \right]^{-1}
\end{equation}
and $\epsilon' = \sqrt{\tau(1+\tau)(1-\epsilon^2)}$ with $\tau = Q^2/4M^2$.  The proton vector-coupling form factor with the $Z$-boson  is given by 
\begin{equation}
    G_{E,M}^{p\mathrm{Z}} = (1- 4\sin^2\theta_W) G_{E,M}^{p\gamma} - G_{E,M}^{n\gamma} - G_{E,M}^{s\gamma}
\end{equation}
and in the limit of vanishing $Q^2$ the $Z$ electric form factor reduces to $1- 4\sin^2\theta_W$ or colloquially the ``weak charge'' of the proton.  The axial charge is related to the isovector charge found in neutron $\beta$ decay under the assumption of SU(3) symmetry, often using input from processes such as hyperon decays.  The SU(3) axial components are related to the spin structure measured in deep inelastic scattering processes \citep{RevModPhys.77.1257}
\begin{equation}
    \Gamma_1^p = \frac{1}{2} \int_0^1 \sum e^2_i \Delta q_i(x) dx \sim \frac{1}{12} g_A^{(3)}  + \frac{1}{36} g_A^{(8)} + \frac{1}{9} g_A^{(0)}  + ...
\end{equation}
with $\Delta q$ the deep inelastic spin structure functions, $e_i$ the quark electric charges, and higher order QCD corrections represented by the ellipses.  This quantity is critical to evaluating the spin contributions of quarks of the nucleon and is an important area of study in hadronic physics.  The axial form factor suffers from poorly-constrained radiative corrections \citep{PhysRevD.72.073003}
\begin{equation}
    G_A^p(Q^2 = 0) = g_A^{(3)}\left(1 + R_A^{T=1}\right) + \frac{3F-D}{2} R^{T=0}_A + \Delta s\left( 1 + R_A^{(0)} \right)
\end{equation}
where $\Delta s = g_A^{(8)} - g_A^{(0)}$, $R_A^T$ are the isovector and isoscalar radiative corrections and $R_A^{(0)}$ is the isosinglet.  In particular the isovector and isoscalar components can have large multiquark (anapole) corrections which give an overall uncertainty of as much as 30\% to the value in this channel.  The parity-violating asymmetry differences between positron and electrons would provide useful data, but would likely take decades of running in a configuration similar to G0 backwards running for meaningful constraint.  The radiative corrections for the asymmetry difference in principle can have different cancellations~\cite{PhysRevLett.94.212301} which would need to be studied for this channel.

The charge asymmetry is dominated by two-photon effects which are orders of magnitude larger, difficult to calculate or predict, and objects of their own study.  It may be possible to go to sufficiently low momentum transfer where the nucleon structure properties are unimportant \cite{PhysRev.74.1759} but the asymmetry may become unobservably small.

\subsection{Deep Inelastic Nucleon Scattering}
Parity violating deep inelastic scattering on fixed targets offers information on the longitudinal momentum quark distributions with new effective couplings complementary to those obtained by electromagnetic processes.  In addition, due to the fact that quarks themselves are highly constrained to be point-like, the scattering rates are much more favorable for high energies.  The parity-violating asymmetry for electron scattering is given by
\begin{equation}
    A_\mathrm{PV} \approx \frac{G_F Q^2}{4\sqrt{2} \pi \alpha} \left[  { a_1(x)} +  \frac{1-(1-y)^2}{1+(1-y)^2} {a_3(x)}  \right] 
\end{equation}
with $y = 1-\frac{E'}{E}$, $E'$ the final lepton energy, $E$ the lepton beam energy and
\begin{equation}
    a_1 (x)  = 2 \frac{\sum C_{1q} e_q ( q+\bar{q})}{\sum e^2_q  ( q+\bar{q} ) }
\end{equation}
\begin{equation}
    {a_3 (x)}  = 2 \frac{\sum {C_{2q}} e_q ({q-\bar{q}})}{\sum e^2_q (q+\bar{q})} 
\end{equation}
with $q$ the quark parton distribution functions, $C_{1,2q}$ the effective quark couplings in the interference 
\begin{equation}
    C_{1q} = \mp\frac{1}{2} + 2 e_q \sin^2\theta_W
\end{equation}
\begin{equation}
    C_{2q} = \mp \frac{1}{2} \pm  2\sin^2\theta_W
\end{equation}
where the upper sign is taken for $u$-type quarks and lower sign for $d$-type quarks, $q$ representing the quark flavor, and $e_q$ the quark electric charge.  Again, the parity-violating asymmetry difference offers direct access to the axial-electron  vector-quark coupling without the requirement of kinematic separation from the larger forward term.  Of note are that the $C_2$ terms contain the differences between the quark and anti-quark distributions, offering an enticing channel of exploration.  A relative figure of merit that is three orders of magnitude worse requires useful measurements of this quantity prohibatively long.

The charge asymmetry difference provides access to a new axial-axial coupling termed $C_{3q}$
\begin{equation}
 \frac{\sigma^+ - \sigma^{-}}{\sigma^+ + \sigma^{-}} = \frac{G_F Q^2}{4 \sqrt{2} \alpha \pi} \frac{1-(1-y)^2}{1+(1-y)^2}  \frac{\sum  C_{3q}  e_q( q - \bar{q} )}{\sum e_q^2 (q + \bar{q})}
\end{equation}
with $C_{3q} = \pm \frac{1}{2}$ with the upper sign taken for $u$-type quarks and lower sign for $d$-type quarks.  These are also proportional to the quark-antiquark differences.  These have been measured once at CERN using $\mu^+$ and $\mu^-$ on carbon nuclei to a relative precision of about 25\% and were in agreement with the standard model prediction \citep{ARGENTO1983245}.

For 12~GeV Jefferson Lab, the asymmetries can be much larger than any parity-violating experiment performed there before.  For the proposed SoLID PVDIS experiment \citep{solid} they are on the order of $10^{-4}$ and for the charge asymmetry, would need to have charge normalization controlled to a level much better than this.  A few percent measurement of $2C_{3u}-C_{3d}$ would take approximately six months of running with $1~\mathrm{\mu A}$ of sufficiently controlled positron-electron running. 

An evaluation of neutral current asymmetries for various electron-ion collider configurations was recently performed~\citep{Zhao:2016rfu}.  The asymmetries may become as large as a few percent and real opportunities for measurements may exist.  The $q-\bar{q}$ extraction, or backwards component, could be improved by an order of magnitude given equal positron integrated luminosity through the charge asymmetry due to circumventing the electron vector coupling suppression. 

\subsection{Fixed Target Electron-Positron Annihilation}

High energy polarized positrons on an electron target would offer a unique opportunity to study low energy parity violation in pair produced leptons.  In terms of standard model interactions, Bhabha scattering at tree level provides identical information to M{\o}ller scattering.  However for sufficient center of mass energies, a fixed target electron-positron annihilation experiment could provide information on electroweak couplings to heavier leptons.  For muon production with an electron beam of energy $E_b$, $s = \sqrt{2 m_e E_b} \ge 2 m_\mu$ or $E_b \ge 43.7~\mathrm{GeV}$, which is outside the reach of the accelerators under considering.

However, if such a machine were to be constructed, it offers an interesting option complementary to the electroweak programs near the $Z$ pole, such as at LEP at CERN or SLD at SLAC.  Here as with the other low energy parity violating programs, the luminosity available with fixed targets is exploited to overcome the intrinsically small observables.  A polarized positron beam could be used to form a left-right forwards-backwards asymmetry of muon pair production that is sensitive to the ratio of the vector to axial coupling ratio \citep{BLONDEL1988438}
\begin{equation}
    A_\mathrm{LRFB} = \frac{(\sigma_\mathrm{LF} - \sigma_\mathrm{RF}) - (\sigma_\mathrm{LB} - \sigma_\mathrm{RB})}{ \sigma_\mathrm{LF} + \sigma_\mathrm{RF} + \sigma_\mathrm{LB} + \sigma_\mathrm{RB} }  \sim \frac{ g_V/g_A}{1+(g_V/g_A)^2}
\end{equation}
where forwards (F) and backwards (B) are integrated over the $\theta < \pi/2$ and $\theta> \pi/2$ hemispheres in the polar angle $\theta$ center-of-mass frame.

The muon the vector coupling has uncertainties an order of magnitude larger relative to the electron \citep{Z-Pole}.  If lepton universality is violated, such as in some of the beyond the standard model explanations for the muon $g$$-$$2$ anomaly \citep{Bennett:2006fi} and proton radius puzzle \citep{doi:10.1146/annurev-nucl-102212-170627}, better measurements of this coupling would be critical in constraining new physics.

\subsection{Parity-violating Elastic Nuclear Scattering from Lead with Coulomb Distortions}

Coulomb distortion effects are critical when studying parity-violating processes on high-$Z$ nuclei, such as in the PREX experiment \citep{PhysRevC.63.025501} which measures the neutron skin thickness of lead-208.  Such an experiment optimizes the sensitivity to the skin thickness at a momentum transfer slightly below the first diffraction peak, the position of which is modified by these distortions.  Prior studies of the lead form factors using elastic electron and positron scattering have confirmed that the diffraction minima for positrons are at larger angle relative to electron scattering~\cite{PhysRevLett.66.572}.  The ability to perform such an experiment at larger angle has many technical advantages over the proposed PREX configuration using the high resolution spectrometers in Hall A at Jefferson Lab.  A cursory evaluation of the Coulomb distortion effects on these measurements using electrons or positrons showed such a difference was likely negligable for the proposed measurements at small angles \citep{horowitz}.

\section{Summary}

While a positron beam offers very interesting possibilities for neutral weak current studies and access to new information complementary to the existing electron programs, the proposed positron beam current and polarization make such measurements extremely challenging.  Further, the charge asymmetry measurements have asymmetries which are of similar magnitude to the parity-violating analogs and would require a charge normalization of incredible precision which is outside of technical capabilities.  Even if such limitations were overcome, a host of beam-related systematic effects, such as control over the beam properties between charge or helicity states, have not been evaluated and would represent an incredible experimental effort requiring many years of dedicated work.  Of the experiments presented, an electron (positron)-ion collider offers the best opportunities for the nearest term progress due to the fact it is at the highest energies and offers the largest asymmetries.

\section{ACKNOWLEDGMENTS}
The author is grateful for productive and stimulating discussions with C.~J.~Horowitz, P.~A.~Souder, and K.~S.~Kumar over the different possibilities of this program.  This work was supported by the U.S. Department of Energy, Office of Science, Office of Nuclear Physics, under contract number DE-AC02-06CH11357.

\nocite{*}
\bibliographystyle{aipnum-cp}%
\bibliography{riordan_jpos17_proc}%

\end{document}